\documentstyle[12pt,epsfig,axodraw]{article}

\newcommand{\dx}{\mbox{\rm d}}
\newcommand{\ra}{\rightarrow}

\newcommand{\s}{\\ \vspace*{-3.5mm} }
\newcommand{\nn}{\noindent}
 
\newcommand{\non}{\nonumber}
\newcommand{\beq}{\begin{eqnarray}}
\newcommand{\eeq}{\end{eqnarray}}
\newcommand{\tb}{\tan\beta}
\newcommand{\mg}{\mu_{\tilde{g}} } 
\newcommand{\mb}{\mu_{\tilde{b}} } 

\begin{document}
\def\thefootnote{\fnsymbol{footnote}}

\begin{flushright}
PM/00--24\\
July 2000
\end{flushright}

\vspace{1cm}

\begin{center}

{\large\sc {\bf Three--Body Decays of SUSY Particles}}

\vspace{0.9cm}

{\sc A. Djouadi$^1$ and Y. Mambrini$^2$} 

\vspace{0.8cm}

Laboratoire de Physique Math\'ematique et Th\'eorique, UMR5825--CNRS,\\
Universit\'e de Montpellier II, F--34095 Montpellier Cedex 5, France. 

\vspace*{.4cm} 

$^1$ djouadi@lpm.univ-montp2.fr \\
$^2$ yann@lpm.univ-montp2.fr \\

\end{center}

\vspace{2.5cm}

\begin{abstract}
\nn We analyze the decays of charginos, neutralinos, gluinos and the
first/second generation squarks in the Minimal Supersymmetric extension of the
Standard Model,  focusing on the three--body decays in scenarios where the
ratio $\tan \beta$ of vacuum expectation values of the two Higgs doublet fields
is large. We show that the three--body decays of the next--to--lightest
neutralinos (lightest charginos) into $b\bar{b},\tau^+\tau^-$ ($\tau \nu$) final
states, where third generation sfermion and Higgs boson exchange diagrams play
an important role, are dominant. Furthermore, we show that decays of gluinos
into $b\bar{b}$ final states and squark decays into lighter sbottoms through
gluino exchange can also have sizeable branching fractions, especially in
scenarios where the soft SUSY breaking gaugino mass parameters are not unified
at the GUT scale.
\end{abstract}

\def\thefootnote{\arabic{footnote}}
\setcounter{footnote}{0}

\newpage

\subsection*{1. Introduction}

Supersymmetric theories (SUSY) \cite{R1} are the best motivated extensions of
the Standard Model (SM) of the strong and electroweak interactions and the
search for supersymmetric particles is one of the major goals of present and
future collider experiments. The Minimal Supersymmetric extension of the
Standard Model (MSSM), with R--parity conservation [leading to a stable
lightest SUSY particle (LSP) which is in general the lightest neutralino
$\chi_1^0$], the minimal particle content [three generations of fermions, two
Higgs doublet fields to break the electroweak symmetry, as well as their SUSY
partners] and a set of soft terms to break SUSY [e.g. mass terms for gauginos,
sfermions and Higgs bosons, and trilinear couplings between squarks and Higgs
bosons], is the most studied in the literature \cite{R1}. \s

In the MSSM, the phenomenology of the third generation sfermions is rather
special. Indeed, due to the large value of the top, bottom and tau lepton
masses, the two current eigenstates $\tilde{f}_L$ and $\tilde{f}_R$ (with
$\tilde{f}=\tilde{t}, \tilde{b}$ or $\tilde{\tau}$) could strongly mix
\cite{qmix}, leading to a large splitting between the two mass eigenstates
$\tilde{f}_1$ and $\tilde{f}_2$ with the lighter one, $\tilde{f}_1$, possibly
much lighter than the other sfermions. In particular, for large values of the
parameter $\tan \beta$, the ratio of the vacuum expectation values of the two
Higgs fields\footnote{Note that large values of the parameter $\tb$, $\tb \sim
50$, are favored in models with unification  of the Yukawa couplings; see e.g. 
Ref.~\cite{btaup}.  [The other solution, with $\tan \beta \sim 1.5$, seems to
be excluded from the negative searches \cite{LEP2} of MSSM Higgs bosons at
LEP2.].}, and non--zero values of the Higgs--higgsino mass parameter $\mu$, the
mixing can be extremely strong in the sbottom and stau sectors, leading to
relatively light $\tilde{b}_1$ and $\tilde{\tau}_1$ states\footnote{Lighter
third generation sfermions have been also advocated in many models such as
those with inverted mass hierarchy where the first and second generation
squarks are much heavier than their third generation partners; for recent
papers see e.g.  Ref.~\cite{inverted}.}.  [In turn, for large values of the
trilinear coupling $A_t$ and/or small $\tb$ values, the mixing can be strong in
the stop sector, leading to a $\tilde{t}_1$ eigenstate much lighter than the
other squarks and possibly lighter than the top quark itself.] \s

A light $\tilde{b}_1$ [and/or $\tilde{\tau}_1$] eigenstate may lead to dramatic
consequences on the decay modes of the next--to--lightest neutralino, the
lightest chargino, the gluino as well as the first and second generation
squarks. Indeed, in this case, one has the following scenarii: \s

$(i)$ The possibility that squarks of the first and second generations decay 
into the lighter $\tilde{b}_1$ squark through the mode 
\beq 
\tilde{q}_{L,R} \ \to \ q \; \tilde{g}^* \ \to \ q \; b \; \tilde{b}_1^* \ + \
q \; \bar{b} \; \tilde{b}_1 
\eeq
opens up. Because it is a QCD mediated decay, and because the decays into light
charginos and neutralinos could be suppressed [e.g., if these particles are
higgsino--like since the couplings are proportional to the light quark mass in
this case], this mode might have a sizeable branching fraction, in particular
in models where the gaugino masses are not unified at the GUT scale. \s

$(ii)$ If gluinos are lighter than squarks, they will mainly decay through 
virtual squark exchange into quarks and charginos/neutralinos; if $\tilde{b}_1$ 
is the lightest squark, its virtuality will be smaller, leading to the possible 
dominance of the decay mode 
\beq
\tilde{g} \ \to \ b \; \tilde{b}^* \; , \; \bar{b} \; \tilde{b} \ \to  \
b \; \bar{b} \; \chi^0_j \ {\rm and/or} \ b \; t \; \chi^\pm_j 
\eeq

$(iii)$ The lightest chargino $\chi_1^+$ and the next--to--lightest neutralino 
$\chi_2^0$ will decay into the LSP and two light fermions. This occurs through
gauge boson, Higgs boson and sfermion exchange diagrams. Because for high $\tb$ 
values, the third generation sfermions, sbottoms and staus, are lighter and the 
Higgs bosons couple strongly to bottom quarks and tau leptons, the three--body 
decays:
\beq
\chi_2^0 & \to &  \chi_1^0 \; \bar{b} \; b \ , \ \chi_1^0 \; \tau^+ \; \tau^- \\
\chi_1^+ & \to &  \chi_1^0 \; \tau^+  \; \nu 
\eeq
are in general enhanced compared to decays where first and second generation 
fermions are involved in the final states. For large values of $\tan \beta$, 
these three--body decay modes have been discussed in Ref.~\cite{Baer} and 
partly in Ref.~\cite{Bartl} in the case of the neutralino $\chi_2^0$. \s

In this note, we analyze these various decay modes and investigate their
consequences. In the next section we discuss the three body decays of squarks
and gluinos, after presenting the analytical expressions for the partial widths.
In Section 3, we analyze the three--body decays of the lightest chargino
$\chi_1^+$ and the second lightest neutralino $\chi_2^0$; we give some
illustrations on the branching ratios of these decay modes and compare our
results with those of Ref.~\cite{Baer}. Some conclusions are then given 
in  section 4.  
 
\subsection*{2. Three--body decays of squarks and gluinos}

The Feynman diagrams for the three--body decays of squarks of the first and 
second generations into the lighter $\tilde{b}_1$ squark, eq.(1), and of 
gluinos through sbottom exchange, eq.(2), are given in Fig.~1. 
 
\begin{center}
\begin{picture}(300,170)(0,0)
\Text(0,120)[]{$\tilde{q}_i$}
\DashArrowLine(10,120)(40,120){4}{}
\ArrowLine(40,120)(70,150)
\Text(77,150)[]{$q$}
\ArrowLine(40,120)(60,90)
\Text(35,100)[]{$\tilde{g}$}
\ArrowLine(60,90)(90,120)
\Text(100,120)[]{$\bar{b}(b)$}
\Text(100,65)[]{$\tilde{b}_1(\tilde{b}_1^*)$}
\DashArrowLine(80,65)(60,90){4}{}
%%%%%%%%
\Text(150,120)[]{$\tilde{g}$}
\ArrowLine(160,120)(190,120)
\ArrowLine(190,120)(220,150)
\Text(234,150)[]{$\bar{b}(b)$}
\DashArrowLine(190,120)(210,90){4}{}
\Text(180,100)[]{$\tilde{b}_1(\tilde{b}_1^*)$}
\ArrowLine(210,90)(240,120)
\Text(250,120)[]{$\chi^0_j$}
\ArrowLine(230,65)(210,90)
\Text(243,65)[]{$b (\bar{b})$}
\Text(150,40)[]{Figure: Generic Feynman diagrams for the three--body decays of 
squarks and gluinos.} 
\end{picture}
\end{center}
\vspace*{-1.3cm}

The Dalitz plot density of the decay mode eq.~(1) is given in terms of the
reduced energies of the two final state quarks, $x_1=2E_q/m_{\tilde{q}_i}$ and
$x_2=2E_b/m_{\tilde{q}_i}$, and the gluino and sbottom reduced squared masses,
$\mu_{\tilde{g}}= m_{\tilde{g}}^2/m^2_{\tilde{q}_i}$ and $\mu_{\tilde{b}}=
m_{\tilde{b}}^2/m^2_{\tilde{q}_i}$. Neglecting the masses of the final state
quarks\footnote{The expressions with the mass effects in this case and for
gluino decays to be discussed later [with the additional final states
$\tilde{g} \to b t \chi_j^\pm$ etc..] are slightly more involved and will be
given elsewhere \cite{these}.}, and adding incoherently the contribution from
the two different $\bar{b} \tilde{b}_1$ and $b\tilde{b}^*_1$ final states, it
is given by: 
\beq
\frac{ \dx \Gamma (\tilde{q}_i) } {\dx x_1 \dx x_2}  &= & \frac{\alpha_s^2} 
{3\pi} \, m_{\tilde{q}_i}  \, \frac{1}{(1-x_1 - \mg)^2} \Bigg[ a_i^2 [(1-x_1+ 
\mb) (1-x_2+\mb)  \non \\ 
&& \hspace*{1cm} +\mb(1- x_1 -x_2 -\mb)]- b_i^2 \mg (1-x_1 -x_2 - \mb) \Bigg] 
\eeq
where the factors $a_i$ and $b_i$ for the squark $\tilde{q}_i$ are given, in 
terms of the $\tilde{b}$ mixing angle $\theta_b$,  by:
\beq
a_1= b_2 = \sin \theta_{b} \ \ {\rm and} \ \ a_2 = b_1= \cos \theta_{b}  
\eeq
Integrating over the energies $x_1,x_2$ with boundary conditions 
\beq
0 \leq x_1 \leq 1- \mu_{\tilde{b}} \ \ {\rm  and} \ \ 
1- x_1 - \mb  \leq x_2 \leq 1- \mb /(1- x_1)
\eeq
one obtains the partial decay width: 
\beq
\Gamma (\tilde{q}_i) &=& \frac{\alpha_s^2} {3\pi} \, m_{\tilde{q}_i} \, 
\Bigg\{ a_i^2 \Bigg[ \frac{(1-\mu_{\tilde g})(\mb -
\mu_{\tilde g})} {2\mu_{\tilde g}^2} (\mu_{\tilde g}-3\mu_{\tilde g}^2
+\mb +\mu_{\tilde g}\mb)
{\rm Log}\frac{\mu_{\tilde g}-\mb}{\mu_{\tilde g}-1} \non \\
&& -\frac{\mb^2}{2\mu_{\tilde g}^2} {\rm Log}\mb
+\frac{\mb-1}{4\mu_{\tilde g}} (5\mu_{\tilde g}-6\mu_{\tilde g}^2
-2\mb +5\mu_{\tilde g} \mb) \Bigg] \non \\
&& \hspace*{1cm} + \ b_i^2 \Bigg[ \frac{(1-\mu_{\tilde g})(\mb -\mu_{\tilde g})}
{\mu_{\tilde g}^2} (\mb-\mu_{\tilde g}^2)
{\rm Log}\frac{\mu_{\tilde g}-\mb}{\mu_{\tilde g}-1} \non \\
&& +\frac{\mb}{\mu_{\tilde g}^2}
(\mu_{\tilde g}-\mb
+\mu_{\tilde g}\mb){\rm Log}\mb
+\frac{\mb-1}{2\mu_{\tilde g}}
(\mu_{\tilde g}-2\mu_{\tilde g}^2
-2\mb+\mu_{\tilde g}\mb) \Bigg] \Bigg\} 
\eeq
%%%%%%%%%%%%%%%%%%%%%%%%%%%%%%%%%%%%%%%%%%%%%%%%%%%%%%%%%%%%%%%%%%%

Neglecting again the mass of the final bottom quark [but not in the couplings],
the Dalitz plot density for the gluino decay $\tilde{g} \to b\bar{b} \chi_1^0$
through $\tilde{b}_i =\tilde{b}_1, \tilde{b}_2$  exchange is simply given by
[here, $x_1=2E_b/m_{\tilde{g}}, x_2= 2E_{\bar{b}}/m_{\tilde{g}}$ and
$\mu_{\tilde{b_i}}= m_{\tilde{b}_i}^2 /m^2_{\tilde{g}}, \mu_{\chi}=
m_{\chi_1^0}^2/m^2_{\tilde{g}}$] 
\beq
\frac{ \dx \Gamma (\tilde{g}) } {\dx x_1 \dx x_2}  &= & \frac{ \alpha_s \alpha}
{4\pi} \;  m_{\tilde g} \;  \sum_{i} (a_{ji}^2+b_{ji}^2) \;  \frac{x_1 
\; (1-\mu_{\chi}-x_1)}{(1-x_1-\mu_{\tilde b_i})^2}
\eeq 
and after integration on the variables $x_1$ and $x_2$, one obtains for the 
partial width: 
\beq
\Gamma(\tilde g)&=&
\frac{ \alpha_s \alpha }{4\pi} m_{\tilde g} \sum_{i} \frac{(a_{ji}^2+b_{ji}^2)}
{\mu_{\tilde b_i}^2} \Bigg\{ \frac{1}{2} \mu_{\tilde b_i} (\mu_\chi -1) 
(5\mu_{\tilde b_i}- 6\mu_{\tilde b_i}^2 - 2\mu _{\chi}+5\mu_{\tilde b_i}
\mu_{\chi} ) \nonumber \\
&+& (\mu_{\tilde b_i}-1)(\mu_\chi - \mu_{\tilde b_i}) (\mu_\chi - 3 \mu_{\tilde 
b_i}^2 + \mu_{\tilde b_i} + \mu_{\tilde b_i} \mu_{\chi}) 
{\rm Log} \frac{1-\mu_{\tilde b_i}}{\mu_{\chi}-\mu_{\tilde b_i}}
-\mu_{\chi}^2 {\rm Log}\mu_{\chi} \Bigg \}
\eeq
where the couplings between neutralino $\chi_j^0$, bottom and sbottoms 
$\tilde{b}_1, \tilde{b}_2$, are given by    
\begin{eqnarray}
\left\{ 
\begin{array}{l}
a_{j1} \\ a_{j2}
\end{array}
\right\}
 &=&  \left\{
\begin{array}{l} c_{\theta_b} \\ -s_{\theta_b} \end{array}
\right\} \Bigg[ \frac{\sqrt{2}}{3}s_W Z'_{j1}-\frac{\sqrt{2}}
{c_W}(-\frac{1}{2}+\frac{1}{3} s^2_W) Z'_{j2}  \Bigg]  
- \left\{ \begin{array}{l} s_{\theta_b} \\ c_{\theta_b} \end{array}
\right\} \frac{m_b}{ \sqrt{2}M_W \cos \beta} Z_{j3}  \nonumber \\ 
\left\{ \begin{array}{l}
b^b_{j1} \\ b^b_{j2}
\end{array}
\right\} &=& \left\{
\begin{array}{l} -c_{\theta_b} \\ s_{\theta_b} \end{array} 
\right\} \frac{m_b}{ \sqrt{2}M_W \cos \beta} Z_{j3} - \left\{
\begin{array}{l} s_{\theta_b} \\ c_{\theta_b} \end{array} 
\right\}  \frac{\sqrt{2}}{3} s_W (Z'_{j1}-\tan \theta_W Z'_{j2}) 
\end{eqnarray}
with $s_W^2=1-c_W^2 \equiv \sin^2\theta_W$ and $s_{\theta_b} \equiv \sin 
\theta_b$, etc..; the (rotated) matrix elements of the neutralino mass matrix 
$Z_{ij}'$ are given in Ref.~\cite{HG}. Note that for massless $b$ quarks, 
there is no interference between the amplitudes with $\tilde{b}_1$ and 
$\tilde{b}_2$ exchange. \smallskip

For illustration we analyze the branching ratios in two models,
discussed in Ref.~\cite{non-universal}, where the gaugino masses are not 
unified at the GUT scale: $i)$ Models in which SUSY breaking occurs 
via an F--term that is not SU(5) singlet but belongs to a 
representation which appears in the symmetric product of two adjoints:
({\bf 24}$\otimes${\bf 24})$_{\rm sym}$={\bf 1}$\oplus${\bf 24}$\oplus${\bf 
75}$\oplus${\bf 200} (where only {\bf 1} leads to universal masses); $ii)$ The 
{\bf OII} model which is superstring motivated and where the SUSY breaking is 
moduli--dominated. The relative gaugino masses at the GUT scale and at the 
scale $M_Z$ are given in Tab.~1 taken from Ref.~\cite{non-universal}. \s

\begin{center}
\begin{tabular}{|c||c|c|c|} \hline
$\ \ \ F_\Phi \ \ \ $ & $\ \ M_3 \ \ $ & $\ \ M_2 \ \ $ & $\ \ M_1\ \ $
\\ \hline
{\bf 1} & $1 (\sim 6)$ & $1(\sim 2)$ & $1 (\sim 1)$ \\ \hline
{\bf 24} & $2 (\sim 12)$ & $-3 (\sim -6)$ & $-1 (\sim -1)$ \\ \hline
{\bf 75} & $1 (\sim 6)$ & $3 (\sim 6)$ & $-5 (\sim -5 )$ \\ \hline
{\bf 200} & $1 (\sim 6)$ & $2 (\sim 4)$ & $10 (\sim 10)$ \\ \hline \hline
{\bf OII} & $1 (\sim 6)$ & $5 (\sim 10)$ & $53/5 (\sim 53/5)$ 
\\ \hline
\end{tabular}
\end{center}
\nn {\small Table 1: Relative gaugino masses at $M_{\rm GUT} (M_Z)$ 
in the $F_\Phi$ representations and the OII model.}

\bigskip 

In Fig.~2, we show the branching ratios of the decays $\tilde{q}_1 \to q b
\tilde{b}_1^*+ q \bar{b} \tilde{b}_1$ (2a) and $\tilde{g} \to \chi^0 b \bar{b}
+\chi^\pm tb$ (2b) as a function of $\mu$ and for $\tb=50$. In Fig.~2a, the
(right--handed) squark mass is taken to be $m_{\tilde{q}_1}= 500$ GeV while the
gluino mass is slightly larger, $m_{\tilde{g}} \sim M_3 = 550$ GeV. For large
values of $\mu$, the lightest neutralinos and chargino are gaugino like, and in
models {\bf 75} and {\bf OII} have masses comparable to the mass of the gluino
$[M_1, M_2, M_3$ are of the same order in this case; the figure is cut in model
{\bf OII} at $\mu \sim M_3$, since the gluino becomes lighter than
$\chi_1^0$]: the two--body squark decays into charginos and neutralinos are
phase space--suppressed and the virtuality of the gluino is not very large,
leading to a rather large BR($\tilde{q} \to q b \tilde{b}_1 $), reaching unity
for the extreme values of $\mu$ where $\tilde{b}_1$ is light.  In models {\bf
1} and {\bf 24}, gluinos are much heavier than the lightest charginos and
neutralinos [$M_3 \gg M_1$], and the two--body decays $\tilde{q} \to q \chi$
dominate [but BR($\tilde{q}_1 \to q b \tilde{b}_1^*$) reaches the level of a
few percent]. Model {\bf 200} is an intermediate case.  For small values of
$\mu$, the lightest chargino and neutralinos are higgsino like and much lighter
than the gluino, $\tilde{b}_1$ is also heavier and the two--body decay
$\tilde{q} \to q \chi$ dominate. \s

In Fig.~2b, the branching ratio $\tilde{g} \to b\bar{b} \chi^0 + bt \chi^\pm$
is shown for $M_{3} \sim m_{\tilde{g}}=350$ GeV and a common squark mass $m_{
\tilde{q}}=600$ GeV; with increasing $\mu$ the lightest sbottom mass varies
from  $\sim 560$ GeV (for $\mu=200$ GeV) to $\sim 357$ GeV (for $\mu=1$ TeV).
Even for the universal gaugino mass case, the branching ratio is very large,
especially for small and large values of $\mu$. For $\mu\sim 200$ GeV,
$\tilde{b}_1$ is not much lighter that the other squarks, but the $\chi^0
\tilde{b} b$ coupling is enhanced [$\propto m_b/M_W \tan \beta$; see eq.~(11)]
and the $b\bar{b}$ final state is favored. For large values of $\mu$,
$m_{\tilde{b}_1}$ becomes significantly smaller than $m_{\tilde{q}}$ and the
sbottom exchange channel dominates leading to the dominance of $\chi^0
b\bar{b}$ and $\chi^\pm tb$ final states. As in the case of squark decays, the
situation is more favorable for bottom quark decays of the gluino in models
{\bf 75} and {\bf OII} [in this case also we have cut the plot at $\mu \sim
M_3$ since then, $m_{\tilde{g}} < m_{\chi_1^0}$], while it is similar for the
{\bf 24} and {\bf 200} models.  For larger $\mu$ values, $\mu \geq 1$ TeV, the
two--body decay $\tilde{g} \to b \tilde{b}_1$ is accessible kinematically and
will of course dominate [since it is a QCD process] all other channels.  

\setcounter{figure}{1}

\begin{figure}[htbp]
\vspace*{-4cm}
\hspace*{-4cm}
\mbox{\psfig{figure=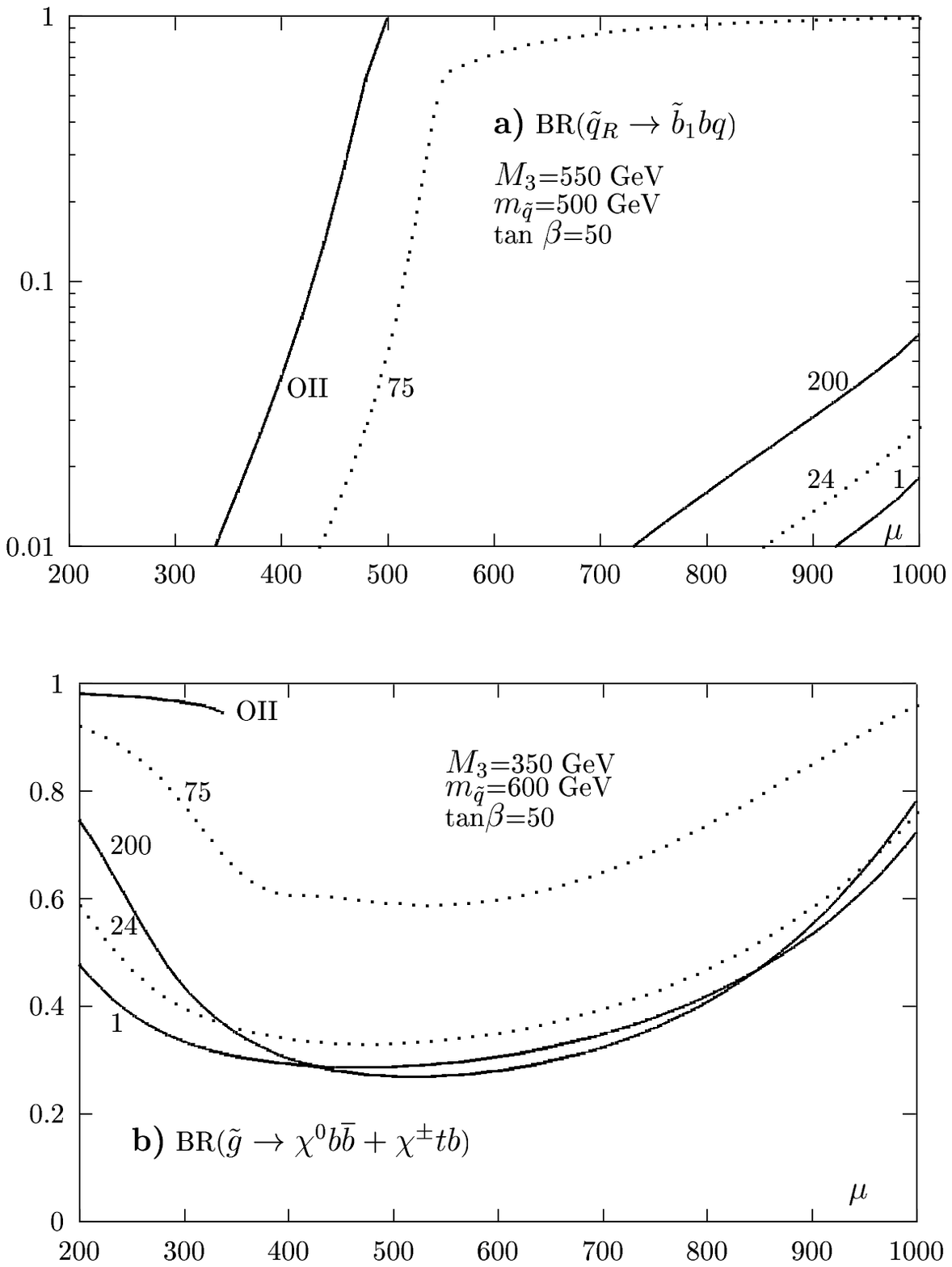,width=22.5cm}}
\vspace*{-9.5cm}
\caption[]{The branching ratios BR$(\tilde{q}_R \to q b \tilde{b}_1^*+ q \bar{b}
\tilde{b})$ for a common scalar squark mass $m_{\tilde{q}}=500$ GeV and 
$M_3=550$ GeV (a) and BR$(\tilde{g} \to \chi^0 b \bar{b} + \chi^\pm tb)$ for 
$m_{\tilde{q}}=600$ GeV and $M_3=350$ GeV (b), as a function of $\mu$ for $\tan
\beta=50$ in the various models discussed above. $M_3$ is fixed and the values 
of $M_1$ and $M_2$ are given Table 1.}  
\end{figure}

\subsection*{3. Chargino and Neutralino Decays} 

We have also calculated the partial decay widths of the three--body decays of
the lightest chargino and of the next--to--lightest neutralino into the LSP and
two fermions, eqs.~(3--4). These decays are mediated by the exchange of
gauge bosons [$W$ boson for $\chi_1^+$ and $Z$ boson for $\chi_2^0$], Higgs
bosons [the charged $H^+$ boson for $\chi_1^+$ and the three neutral Higgs
bosons $h,H$ and $A$ for $\chi_2^0$] as well as $t$ and $u$ channel sfermion
exchanges [in particular the lighter $\tilde{\tau}_1$ slepton for $\chi_1^+$ 
and $\tilde{b}_1, \tilde{\tau}_1$ states for $\chi_2^0$; Fig.~3. \s

\begin{picture}(1000,170)(0,0)
\Text(0,120)[]{$\chi$}
\ArrowLine(10,120)(40,120)
\ArrowLine(40,120)(70,150)
\Text(77,150)[]{$\chi_1^0$}
\Photon(40,120)(60,90){4}{8}
\Text(35,100)[]{$V$}
\ArrowLine(60,90)(90,120)
\Text(100,120)[]{$f$}
\ArrowLine(80,65)(60,90)
\Text(100,65)[]{$\bar{f}$}
%W exchange
\Text(150,120)[]{$\chi$}
\ArrowLine(160,120)(190,120)
\ArrowLine(190,120)(220,150)
\Text(227,150)[]{$\chi_1^0$}
\DashArrowLine(190,120)(210,90){4}
\Text(185,100)[]{$\Phi$}
\ArrowLine(210,90)(240,120)
\Text(250,120)[]{$f$}
\ArrowLine(230,65)(210,90)
\Text(250,65)[]{$\bar{f}$}
%charged Higgs  exchange
\Text(300,120)[]{$\chi$}
\ArrowLine(310,120)(340,120)
\ArrowLine(340,120)(370,150)
\Text(377,150)[]{$f$}
\DashArrowLine(340,120)(360,90){4}{}
\Text(335,100)[]{$\tilde{f}^*$}
\ArrowLine(360,90)(390,120)
\Text(400,120)[]{$\bar{f}$}
\ArrowLine(380,65)(360,90)
\Text(400,65)[]{$\chi_1^0$}
\end{picture}

\vspace*{-1.5cm}

\nn Figure 3: Generic Feynman diagrams contributing to the 
three--body decays of charginos and neutralinos into the LSP and two 
fermions. \vspace*{4mm} 

We have taken into account the contributions of all the channels [and of course
the interference terms], the full dependence on the masses of the final
fermions [to be able to describe more accurately the cases of chargino decays
into $\tau \nu, tb$ and the neutralino decays into $b\bar{b}, t\bar{t}$ final
states] and the mixing of the third generation sfermions.  Complete analytical
formulae for the Dalitz plot densities have been obtained; for the integrated
partial widths, exact formulae have been derived in the case where the fermions
in the final state are massless [in Refs~\cite{Baer,Bartl}, the fermion mass
effects have not been taken into account and the totally integrated partial
widths have not been derived]. The lengthy and cumbersome analytical
expressions will be found in Ref.~\cite{these}. In this section, we will simply
exhibit the behavior of the $\chi_2^0$  and $\chi_1^+$ branching ratios in the
case of decays into $b\bar{b}$ and $\tau \nu$ final states, respectively [which
are of more immediate interest, being in the range probed at LEP2 and the
Tevatron].  They are shown in Figs.~4 and 5 for two values of $\tan \beta$ and
the pseudoscalar Higgs boson mass $M_A$; $M_2$ is fixed to $M_2=140$ GeV and we
assumed universal gaugino masses at $M_{\rm GUT}$ [this leads for $\mu \sim
450$ GeV and $\tan \beta=50$ to the set of gaugino masses: $m_{\chi_1^0} \sim
60$ GeV and $m_{\chi_2^0} \sim m_{\chi_1^+} \sim 135$ GeV; the variation with
$\tan \beta$ is mild].  The trilinear couplings  $A_i$ are fixed to 100 GeV for
squarks and sleptons.  \s

\setcounter{figure}{3}

\begin{figure}[htbp]
\vspace*{-3.5cm}
\hspace*{-3.8cm}
%\begin{center}
\mbox{\psfig{figure=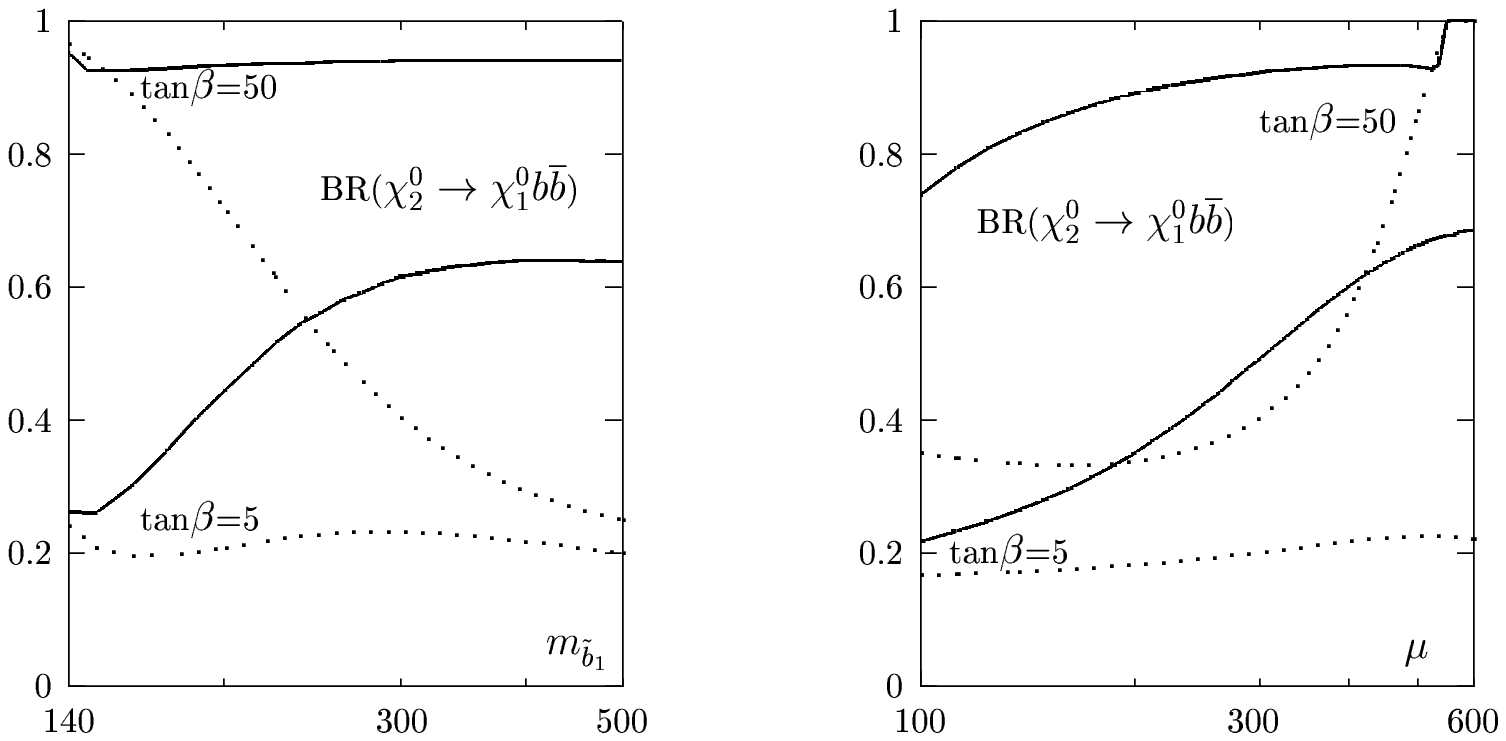,width=22.5cm}}
%\end{center}
\vspace*{-21.cm}
\caption[]{The branching ratio BR$(\chi^0_2 \ra \chi_1^0 b \bar{b})$ as
a function of $m_{\tilde{b}_1}$ for $\mu=450$ GeV (left) and as a function of 
$\mu$ (right) [assuming in this case, a common scalar mass for squarks and 
sleptons, $\tilde{m}= 380$ GeV] for two values of $\tan \beta=5,50$ and $M_A=
100$ GeV (full lines) and 500 GeV (dotted lines); $M_2$ is fixed to $M_2=140$ 
GeV.}  
\end{figure}

\begin{figure}[htbp]
\vspace*{-3.5cm}
\hspace*{-3.8cm}
%\begin{center}
\mbox{\psfig{figure=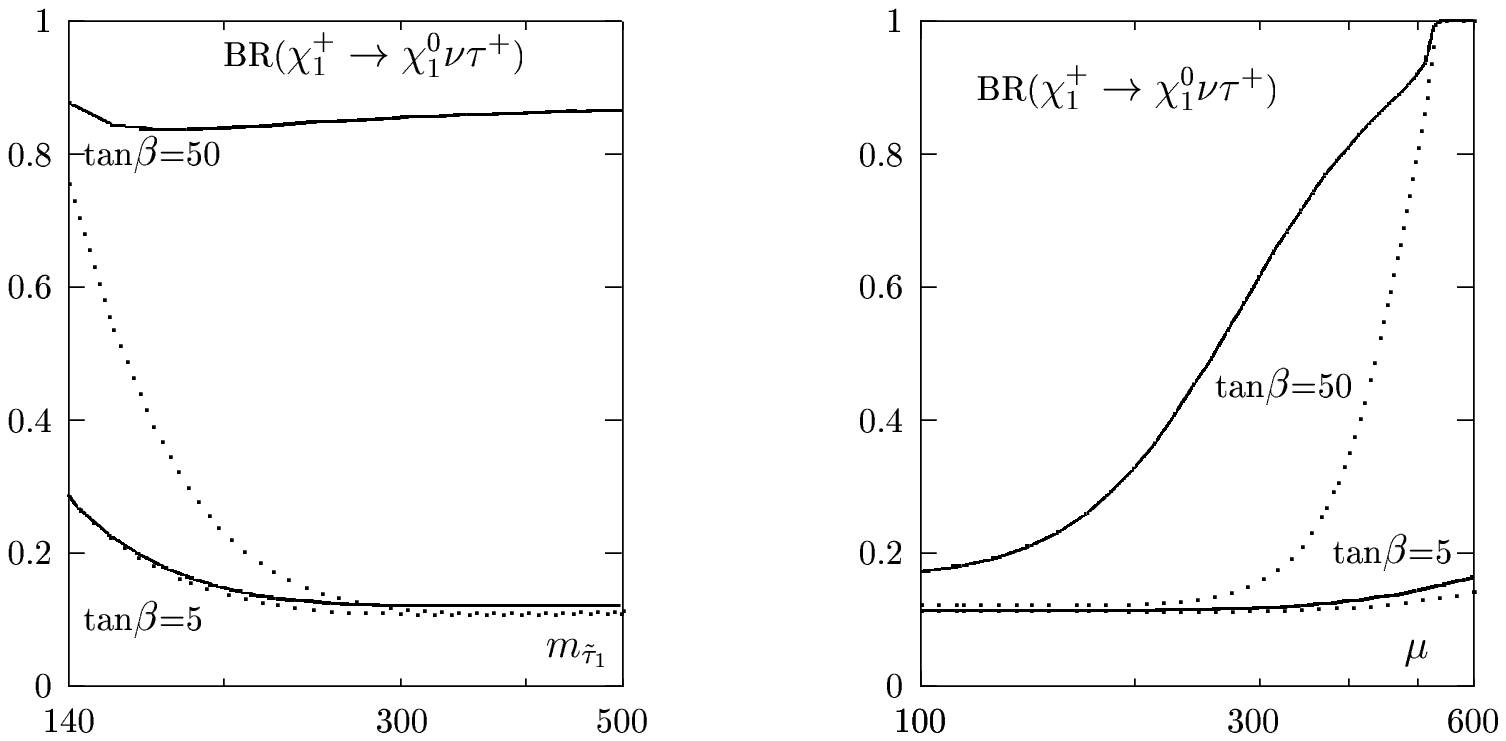,width=22.5cm}}
%\end{center}
\vspace*{-21.cm}
\caption[]{The branching ratio BR$(\chi^+_1 \ra \chi_1^0 \nu \tau^+)$ as
a function of $m_{\tilde{\tau}_1}$ for $\mu=450$ GeV (left) and as a function 
of $\mu$ (right) [assuming, in this case, $\tilde{m_{q}}= 400$ GeV and 
$\tilde{m_{l}}= 250$ GeV] for two values of $\tan \beta=5,50$ and $M_A=100$ 
GeV (full lines) and 500 GeV (dotted lines); $M_2$ is fixed to $M_2=140$ GeV.}  
\end{figure}

In Fig.~4, BR($\chi_2^0 \to \chi_1^0 b\bar{b}$) is shown as a function of
$m_{\tilde{b}_1}$ (left) and $\mu$ (right). For large values of $\tb$, the
$Ab\bar{b}$ and $hb \bar{b}$ couplings are strongly enhanced, and for $M_A=100$
GeV, the Higgs exchange contributions are largely dominant pushing the
branching ratio from $\sim 20\%$ [when only the $Z$--boson exchange
contribution is important] to values close to 90\% [the remaining 10\% are in
general taken by the branching ratio of the decays into $\tau$ lepton pairs]. 
For $M_A=500$ GeV, the Higgs exchange contribution is suppressed but for
relatively small values of $m_{\tilde{b}_1}$ or large values of the parameter
$\mu$ [leading again to a small $m_{\tilde{b}_1}$], BR($\chi_ 2^0 \to \chi_1^0
b \bar{b}$) is still enhanced. Note that even for moderate $\tan \beta$, $\sim
5$, the branching ratio can be large if $M_A$ is small and/or $\mu$ is large.  

Fig.~5 displays the branching ratio for the lightest chargino decay
$\chi_1^+ \to \chi_1^0 \tau^+ \nu_\tau$ as a function of $m_{\tilde{\tau}_1}$
(left) and $\mu$ (right). The figure shows the same pattern as for the previous
decay: for large values of $\tan \beta$, a relatively light charged Higgs
contribution can strongly enhance BR($\chi_1^+ \to \chi_1^0 \tau^+ \nu_\tau$),
but even with a heavier $H^\pm$ boson, the branching ratio can exceed the level
of 80\% because of a lighter $\tilde{\tau}_1$ eigenstate.  \s

We have developed a fortran code called {\tt SDECAY} \cite{sdecay} which
calculates the partial decay widths and branching ratios of the chargino,
neutralino and gluino decays discussed in this note\footnote{The program
contains, in addition, the branching ratios for the four--body decay modes
[which account also for the three--body decays] of the lightest top squark
\cite{4body}.}. For the supersymmetric spectrum [including the renormalisation
group equations for parameter evolution] and for the parameterization of the
Higgs sector, it has been interfaced with the programs {\tt SUSPECT}
\cite{suspect} and {\tt HDECAY} \cite{hdecay}. We have compared our results
with those of Ref.~\cite{Baer} which have been implemented in the program {\tt
ISAJET} \cite{ISAJET}. For massless fermions, the agreement was
perfect\footnote{The comparison was slightly tricky since the evolution of the
couplings and the soft SUSY--breaking terms as well as the parameterization of
the Higgs sectors are given in different approximations in the programs {\tt
SUSPECT} and {\tt ISAJET} and we needed to use the same input parameters at low
energy in both programs. We thank Laurent Duflot from ALEPH for his help with
this comparison.  An independent check in the case of the chargino decays, has
also been performed by F. Boudjema and V. Lafage \cite{FB}.}, giving a great
confidence that this rather involved calculation is correct.

\subsection*{4. Conclusions} 

In this note, we have analyzed the three--body decay modes of gluinos, squarks
and those of the lightest charginos and next--to--lightest neutralinos in the
MSSM. We have made a complete calculation of the decay widths and branching
ratios, taking into account all possible channels, the mixing in the sfermion
sector and the finite masses of the final particles, and provided a fortran
code for the numerical evaluation of the branching ratios. \s

For large values of $\tan \beta$, the bottom and tau Yukawa couplings become
large, leading to smaller masses of the tau slepton and bottom squark compared
to their first and second generation partners.  This leads to enhanced
branching ratios of gluinos into $b\bar{b}$ final states and to the possibility
that squarks decay into lighter sbottoms through gluino exchange can have
sizeable branching fractions; this is particularly the case in scenarios where
the soft SUSY breaking gaugino mass parameters are not unified at the Grand
Unification scale. The branching ratios of the decays of the lightest chargino
into $\tau \nu$ final states and of the next--to--lightest neutralinos into
$b\bar{b}$ and $\tau^+ \tau^-$ pairs are also enhanced in the large $\tb$
scenario, with an additional increase being due to the stronger Yukawa
couplings to charged and neutral Higgs bosons, respectively. \s

Thus, SUSY events will contain more $b$--quarks and $\tau$ leptons in the final
state than initially expected. This renders the search for SUSY particles and
the measurement of the SUSY parameters, where the electron and muon channels
where used, less straightforward as already discussed in Ref.~\cite{Baer2}. 
$b$--tagging and the identification of the decays of the tau leptons become
then a crucial issue in the search and the study of the properties of these
particles, in particular at hadron colliders such as the Tevatron and LHC.  

\bigskip 

\nn {\bf Acknowledgments:} This work is supported in part by the 
GDR--Supersym\'etrie. We thank the members of the GDR, in particular Fawzi 
Boudjema, Daniel Denegri, Laurent Duflot, Jean--Francois Grivaz, 
Stavros Katsanevas, Jean-Loic Kneur, Vincent Lafage and Gilbert Moultaka 
for discussions.

\end{document}